\documentclass[12pt]{article}

\usepackage{times}

\usepackage{graphicx}
\usepackage[numbers,sort&compress]{natbib}
\newcounter{lastnote}
\addtocounter{lastnote}{+1}%

\title{Selective Interface Control of Order Parameters in Complex Oxides }

\author
{D. Meyers$^{1,\ast}$, Jian Liu$^{2,\ast}$, J. W. Freeland$^{3}$, S. Middey$^{1}$, M. Kareev$^{1}$, J. M. Zuo$^4$, \\ Yi-De Chuang$^5$, J.-W. Kim$^{3}$, P. J. Ryan$^{3}$, and J. Chakhalian$^1$\\
\\
\normalsize{$^{1}$Department of Physics, University of Arkansas, Fayetteville, AR 72701, USA}\\
\normalsize{$^{2}$Department of Physics, University of California, Berkeley, CA 94720, USA}\\
\normalsize{$^{3}$Advanced Photon Source, Argonne National Laboratory, Argonne, IL 60439, USA}\\
\normalsize{$^{4}$Department of Materials Science and Engineering, University of Illinois, Urbana, IL 61801, USA}\\
\normalsize{$^{5}$Advanced Light Source, Lawrence Berkeley National Laboratory,
Berkeley, CA 94720, USA}\\
\\
\normalsize{$^\ast$Both authors contributed equally; E-mail: dmeyers@uark.edu,jian.liu@berkeley.edu}
}

\date{}
\usepackage{color}

\begin{document} 


\baselineskip24pt

\maketitle 

\begin{abstract}

In complex materials observed electronic phases and transitions between them  often involves coupling between many degrees of freedom whose entanglement convolutes understanding of the instigating mechanism. Metal-insulator transitions  are one such problem where coupling to the structural, orbital, charge, and magnetic order parameters frequently obscures the underlying physics. Here, we demonstrate a way to unravel this  conundrum by heterostructuring a prototypical  multi-ordered complex oxide NdNiO$_3$ in ultra thin geometry, which  preserves the metal-to-insulator transition and  bulk-like magnetic order parameter, but entirely suppresses the symmetry lowering and charge order parameter. These findings illustrate the utility of heterointerfaces as a powerful method for removing competing order parameters to gain greater insight into the nature of the transition, here revealing that the magnetic order   generates the transition independently, leading to a purely electronic Mott metal-insulator transition.

\end{abstract}

\section*{Introduction}


One of the greatest challenges of condensed matter physics involves exposing the true underlying mechanisms giving rise to the observed anomalous properties, a situation greatly complicated by the coupling of various interactions, for example competing nematic, structural and spin transitions in iron pnictide\cite{Fernandes,Johnston2} or  intertwined charge, magnetic and superconducting order parameters in underdoped high-Tc cuprates\cite{Subir,Arm}. In strongly correlated electronic materials, the notion of complexity has been synonymous with multiple and often antagonistic  ordered phases of intertwined charge, spin, and orbital degrees of freedom\cite{Subir,Arm,Salamon,Medarde,Catalan}. True insight into the ground state of these materials thus necessitates the ability to selectively eliminate these degrees of freedom to reveal individual contributions.

As a classic case in question, the crossover of an electrically conducting state of a solid into a phase wherein the  movement of carriers is prohibited is a prototypical example of such problem. This metal-to-insulator transition (MIT)\ is frequently accompanied by  emergent order parameters including structural modulation, magnetic, charge, and  orbital orderings etc., making it an arduous task to decipher the decisive interaction behind the transition\cite{Imada}. Despite these complications, metal-insulator transitions have been controllably modified by external stimuli in an effort to disentangle the coupled order parameters to  the true progenitor\cite{Qazilbash,Morrison,Stojchevska,Ahn1,Jeong,Bollinger,Limelette,Kuwahara}. Congruent to this effort, a deterministic control over the interfaces between layers with distinct  or competing order parameters has  further widened the traditional modalities that govern the global phase behavior of correlated electrons\cite{Jak_RMP,Mannhart,Hwang,Jak1,Jak2,JakNAT}. The  heterointerface  approach  naturally brings forward the important question of whether it is possible to selectively modulate a specific ordering to reveal the primary cause for the  phase transition into a multi-ordered ground state. Unlike the previously mentioned efforts, where the system comes back to the original ground state when the external stimulus is removed, the present study undertook to suppress order parameters by the virtue of epitaxial stabilization, effectively freezing the system in an atypical state.

Specifically, a 15 unit cell thin film of rare-earth nickelate NdNiO$_{3}$ (NNO) is utilized as a model system exhibiting a first-order MIT that in the bulk involves  structural, charge, and antiferromagnetic order parameters whose entanglement has obscured true understanding of the mechanism underpinning the transition, Fig. 1A and B\cite{Medarde,Catalan,JakPRL,JianPRL,JianNCOM,JianPRB,Mazin,Frano_Spiral,Hepting,Wu_PNOPLO,Hauser,Benc,JianNNO,Bodenthin,Staub,Lorenzo,Scagnoli,Scagnoli2,Yamamoto}. Interestingly, recent work by Hepting and Wu \textit{et al} has shown that superlattices utilizing PrNiO$_3$ and PrAlO$_3$ can supress the MI- and CO transitions, while preserving the AFM transition, leading to a rare metallic AFM state\cite{Hepting,Wu_PNOPLO}. Our experiment, spanning x-ray absorption spectroscopy (XAS) and resonant x-ray scattering (RXS), demonstrates that in the ultra thin limit for films the MIT persists with the same bulk-like E'- antiferromagnetic  ordering and changes in electronic structure while the charge order parameter, and accompanying structural transition,  are completely removed at  all  temperatures.  These findings  imply the exceptional case of an isosymmetric and purely  electronic Mott transition\cite{Limelette,Moore_SCRO} driven by  strongly  correlated  electron moments and is in sharp contrast to the present  understanding of  physics of rare-earth nickelates\cite{Staub,Lorenzo,Scagnoli,Scagnoli2}.

\section*{Electronic and magnetic configuration}

Reduction of the degrees of freedom through heterostructuring presumably alters the electronic structure from it's bulk-like state. Indeed, in nickelates, thin-film geometry and  proximity  to  the interface has been  shown to  strongly alter the electronic structure of the constituent layers and, thus, requires investigation; numerous  XAS reports detail the change of the electronic structure across the MIT  showing  a characteristic splitting of the Ni L$_3$ edge below the MIT into two distinct peaks and a narrowing of the $d$-electron bandwidth in the insulating state\cite{Medarde,Catalan,JianNCOM,JianNNO,DerekPRB,John}.  These two distinct effects are the spectroscopic signatures of the stabilization of an insulating state in the nickelates. As seen in Fig. 2A, in the case of ultra thin films of NNO, the Ni L$_3$ edge does indeed show a clear splitting below the MIT. Tracking the intensity in between the two peaks (inset of Fig. 2A) confirms a distinct spectroscopic change  quantitatively very close to bulk-like behavior across the MIT\cite{John}. Similarly, the O K-edge pre-peak reflects the band narrowing across the MIT, Fig. 2B; the sudden shift in bandwidth is commensurate with the first-order MIT at  $\sim$ 150K,  (Fig. 3B, solid lines)\cite{JianNCOM}. Thus,  both XAS   and transport measurements affirm that the ultra thin structural motif does not generate any anomalous electronic structure effects across the MIT, making it an ideal candidate for investigation of the commensurate order parameters.

Spin ordering is a prevalent ingredient in Mott transitions\cite{Imada,Khomskii}. In the nickelates, the magnetic  ordering  has received a widespread attention due to the unusual stacking of ferromagnetic planes along the (1 1 1)$_{pc}$ (pc = pseudo-cubic) direction that are coupled antiferromagnetically (AFM) to one another in an up-up-down-down pattern, a non-collinear periodic behavior and a magnetic unit cell consisting of four structural unit cells, shown in Fig. 1C\cite{Scagnoli3,Scagnoli4}. Probing this anomalous, E'- AFM ordering in ultra-thin film geometry is quite challenging; Fig. 3A displays the results of the soft x-ray resonant scattering (RXS) at the (1/2 0 1/2)$_{or}$ (or = orthorhombic) reflection with the energy tuned to the Ni L$_3$ edge (852 eV) below the MIT. This structurally  forbidden Bragg reflection corresponds to a 4-fold unit cell repetition in the (1 1 1)$_{pc}$ direction. As  seen in  Fig. 3B, circles, the intensity of this reflection tracks very close with the MIT, suddenly rising above the background noise at around 140K and steadily increasing until beginning to stabilize at low temperature. Both the periodicity and spectroscopic signature are in excellent agreement with previous studies on thick NdNiO$_3$ films and bulk powders\cite{Garcia,Scagnoli3}.  In short, these results  show that the bulk E$^\prime$-type AFM order parameter is preserved and conforms with the MIT  despite bi-axial strain ($\sim$ 1.4 \%). With the expected  electronic structure response (i.e. AFM order parameter, and first-order MIT) the pinning of the lattice to the substrate does not  cause any anomalous perturbation to the bulk-like magnetic and transport behavior of the nickelate film. In addition, for any bulk rare-earth nickelate,\ the ground state is characterised by the  presence of  charge ordering (CO)\ \textit{and}  the structural transition  from the orthorhombic
$Pbnm$ symmetry of the metallic phase  to monoclinic $P2_1/n$ symmetry of the insulating phase \cite{Medarde,Catalan,Lorenzo}. 

\section*{Probing lattice symmetry and charge ordering }

First,  we discuss the issue of lattice symmetry  transformation,  which   is considered to  be critical  for the  
MIT. Heterostructuring naturally leads to a modulation of the film lattice due to the strong bonding with the substrate's ions. When the film becomes thick enough the relaxation of  elastic strain  is inevitable and effectively decouples the film from the substrate\cite{Kaul}. In the ultra thin regime, however, the film is pinned to the substrate with no detectable relaxation and the heteroepitaxy infact  controls the lattice degrees of  freedom therein\cite{JianNNO,DerekPRB}. 

$Pbnm $ (metal)\ and  $P2_1/n$ (insulator)\ space groups  share the same Ni arrangement,  however  they are split into different Wycoff positions with the symmetry lowering. These inequivalent Ni sites carry a rock-salt pattern of charge disproportionation Ni$^{3\pm\delta}$ giving rise to the CO parameter. In recent years it has been found that, while hard RXS is a powerful tool for investigating charge ordering, careful  analysis is required to avoid the misinterpretation of CO  for  small distortions of the oxygen
octahedral  network  \cite{Lorenzo,Staub,Garcia2}. With this caution in mind, we investigated the (0 1 5)$_{or}$ and (1 0 5)$_{or}$ reflections, which are conventionally used to probe the  lowering of the symmetry to monoclinic $P2_1/n$, Fig. 4A and B\cite{Lorenzo,Staub,Garcia2,Footnote1,Upton}.

 Fig. 4A displays scans along the L reciprocal space vector (L-scan) at the (0 1 5)$_{or}$ and (1 0 5)$_{or}$ peaks. The (1 0 5)$_{or}$ peak is symmetry allowed for orthorhombic NNO, as a Bragg peak corresponds to the Nd sublattice, thus the film peak with Kiessig fringes is anticipated. As expected, this (105) reflection should have  no Ni contribution until the charge ordering breaks the \textit{Pbnm} symmetry in the low temperature insulating phase; the CO  then leads to an additional contribution  to the peak from Ni causing a sharp change in signal strength, especially when the x-rays are tuned to the resonant Ni K-edge (8.34 keV)\cite{Lorenzo}. Surprisingly, as the temperature  is scanned across the MIT no detectable change in the peak intensity is observed, Fig. 4A inset. This result immediately implies that neither charge ordering nor the associated symmetry breaking occurs  across the transition. In addition, the (0 1 5)$_{or}$ peak, which is  symmetry forbidden for \textit{Pbnm} , does not appear at any temperature, thus confirming the isosymmetric  nature of the MIT.

Furthermore, a key feature of resonant scattering is that the additional terms within the scattering factor are $highly$ sensitive to the x-ray energy around an absorption edge\cite{Fink,Hodeau}. Fig. 4B shows the energy scan at the (1 0 5)$_{or}$ peak which further corroborate the above picture with higher sensitivity,  confirming that \textit{no Ni resonance signal (i.e. symmetry  lowering)\  is detected below the MIT}. This is in stark contrast to all previous reports on both thick films and bulk where strong, temperature dependent resonance was shown to  track with the MIT\cite{Staub,Lorenzo,Scagnoli,Scagnoli2}.  To further verify this finding, the allowed (2 2 0)$_{or}$ reflection was measured and shows the expected Ni resonance signal, confirming that the Ni contribution is certainly detectable in our experimental setup. These results confirm that across the MIT\ (\textit{i}) no bond disproportionation of NiO$_6$  occurs and the metallic phase \textit{Pbnm} lattice  symmetry is preserved \cite{Lorenzo, Scagnoli}, and (\textit{ii}) since no detectable Ni resonance
is observed no charge ordering occurs emerges in the insulating phase . These findings imply that the ultra thin films have stabilized a previously unknown nickelate ground state consisting of an insulating orthorhombic phase with  AFM order. Intriguingly, as  observed here, the case of a phase transition without a structural symmetry change can only be first order and is exceptionally  rare in complex materials, with the most prominent examples being analogous to the liquid-gas transformation\cite{Khomskii,Shimahara}. For complex oxides, there are only  two known cases of this type of MIT,  i.e. Cr-doped V$_2$O$_3$ \cite{Limelette} and the surface driven Ca$_{1.9}$Sr$_{0.1}$RuO$_4$\cite{Moore_SCRO}.

\section*{Theoretical Presage}

Driven by the heterointerface, CO removal and  the stabilization of the unknown Mott phase within this  class of  materials is of great interest  and yet has some precedent in past theoretical work \cite{Yamamoto,SungBin,Sungbin2,Prosandeev,Mizokawa,Lau,Johnston,Park,Park2}.  For example, two recent studies utilizing different theoretical methods by Lee \textit{et al} \cite{SungBin,Sungbin2} have  proposed that the CO is slaved to the $E^{\prime}$-magnetic ordering in the weak coupling limit, and can indeed disappear under certain conditions; in particular, using Landau theory, the theory  suggests  that restricting the nickelates to the ultra-thin film regime may  remove the CO. On the other hand,  the predicted phase  changes the $Q$-vector for the antiferromagnetic ordering, which  is  in variance with the experiment. Beyond this , Park \textit{et al}  have demonstrated that within dynamical mean field theory (DMFT), despite the near Fermi-energy imbalance in the spectral weight between the two Ni sites, the total valence of Ni on both sites is practically identical, with the two different Ni sites instead hybridizing with O, leading to an S = 1 state on the larger octahedra (3d$^8$) and an S = 0 state formed due to AFM coupling with the O holes (3d$^8$\underline{\underline L})\cite{Park,Park2}. However, when the lattice symmetry  is raised to \textit{Pbnm}  a metallic state with no MIT has emerged. More recently, Johnston \textit{et al} \cite{Johnston} utilized Hartee-Fock methods to show the NiO$_6$ octahedra form a alternating collapsed and expanded octahedra, giving a $d^8$ + $d^8$\underline{L}$^2$ state, where no CO on Ni occurs.

Finally, using LSDA + U calculations, Yamamoto \textit{et al} \cite{Yamamoto} obtained results that are in the good agreement with   our observations. Specifically,  the calculated  electronic and magnetic structure in orthorhombic NNO is found to  be an insulating state with no Ni CO (as expected for equivalent Ni sites  in $Pbnm$ symmetry). In addition, the calculation shows that magnetic space group is lowered to monoclinic due to different spin density polarizations around two O sites that preserve the equivalence of Ni sites in the \textit{Pbnm} space group. Most importantly, this symmetry breaking state involving holes on oxygen and  driven by the Hubbard U, opens an insulating gap, which agree well with the previous work \cite{Stewart}. At  this point we can conjecture that while in the bulk structural symmetry is indeed lowered to \textit{P2$_1$/n,}  the epitaxial interface is  able to preserve the orthorhombic  structural symmetry of the metallic phase. The resulting ground state, observed experimentally, can be obtained  within  the LSDA + U framework, supporting the notion that the bulk-like MIT and magnetic order parameter can be attained with the charge and structural order parameters removed. In this work, we find  the heterointerface acts as a powerful tool to effectively isolate the magnetic order parameter, which drives the bulk-like MIT independently. Thus, by constraining symmetry solutions via substrate imprinting, it is possible to bound order parameter space allowing inter-order behavior to be studied.

\section*{Concluding Remarks}


In conclusion, the reduction of the number of simultaneously  competing order parameters commensurate with a phase transition from a metallic to a Mott insulating state has been achieved on a prototypical  ultra thin film of NNO. The thin film heteroepitaxy prevents   symmetry lowering from  \textit{Pbnm} to \textit{P2$_1$/n} across the MIT, thus removing the bulk-like CO parameter. Despite this anomalous state, the  Mott MIT persists with no significant effect on the magnetic order parameter. The magnetic order parameter is  identified as the culprit  which  drives the pure electronic MIT in the nickelates, highlighting the utility of this emerging method to sunder the competing order parameters. Our findings suggest that application of this method to eliminate  specific order parameters to highly entangled or ``hidden" orders  found in  cuprates, pnictides, heavy fermions and chalcogenides families may  shed  new light on their anomalous ground states.

\section*{Acknowledgments}

DM, JF and JC were supported  by DOE Grant No:~DESC0012375 for synchrotron work at the Advanced Photon Source. SM was supported by the DOD-ARO grant. Synthesis and basic characterization at the University of Arkansas was supported  in part by the Gordon and Betty Moore Foundation EPiQS Initiative through Grant GBMF4534. Work at the Advanced Photon Source is supported by the U.S. Department of Energy, Office of Science under grant No. DEAC02-06CH11357. Work at the Advanced Light Source is supported by the U.S. Department of Energy, Office of Science under Contract No. DE-AC02-05CH11231.

\clearpage

\begin{figure}[t!]\vspace{-0pt}
\includegraphics[width=\textwidth]{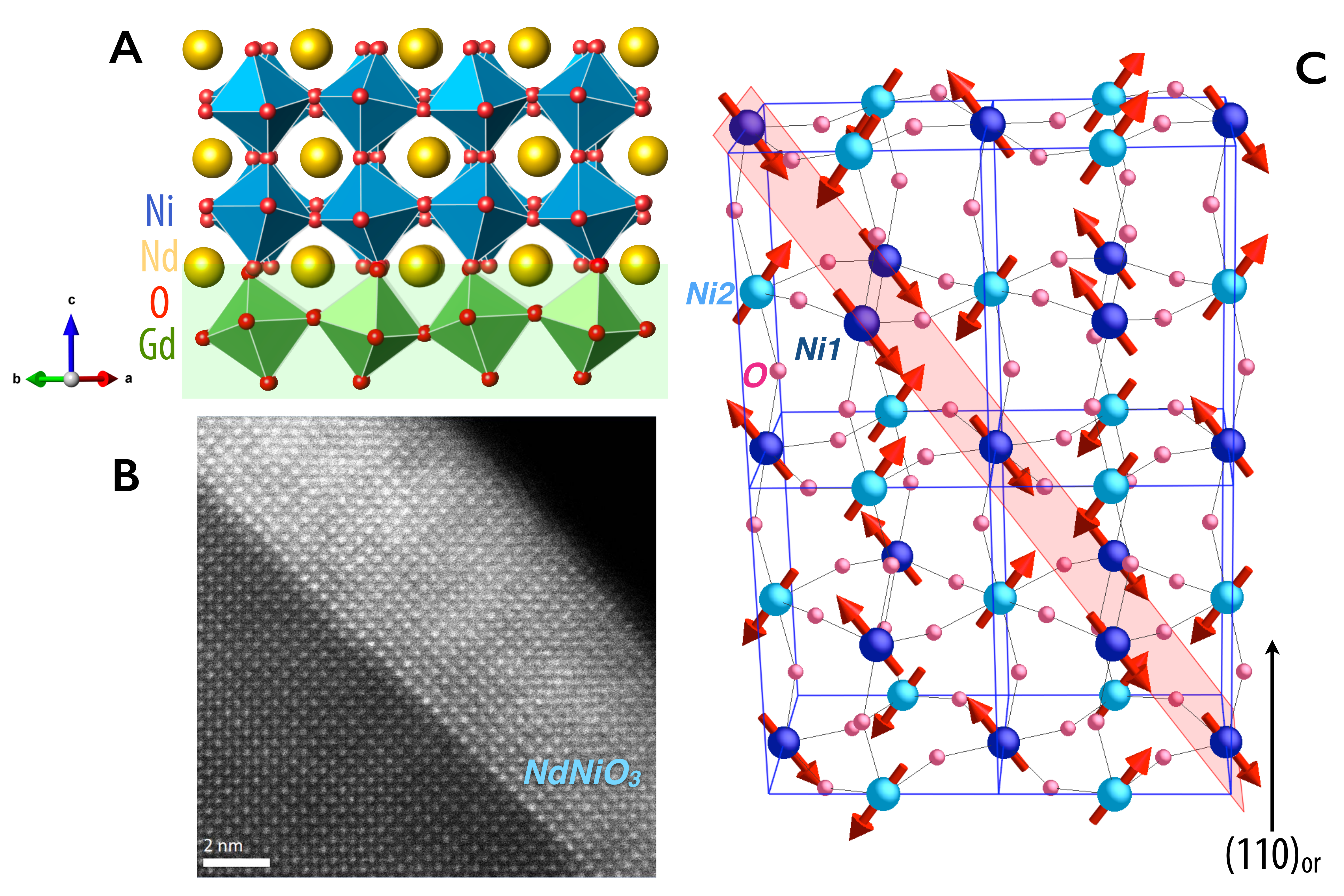}
\caption{\label{} (Color online) (\textbf{A}) Heterostructure interface of NNO grown on NGO. (\textbf{B}) TEM showing atomically sharp interface. (\textbf{C}) E$^\prime$-type antiferromagnetic ordering in the nickelates with the (111)$_{pc}$ plane highlighted. The dark and light blue spheres represent the nickel sites with charge of 3 $\pm$ $\delta$\cite{Scagnoli3,Scagnoli4}.}
\end{figure}

\clearpage

\begin{figure}[h!]\vspace{-0pt}
\includegraphics[width=.8\textwidth]{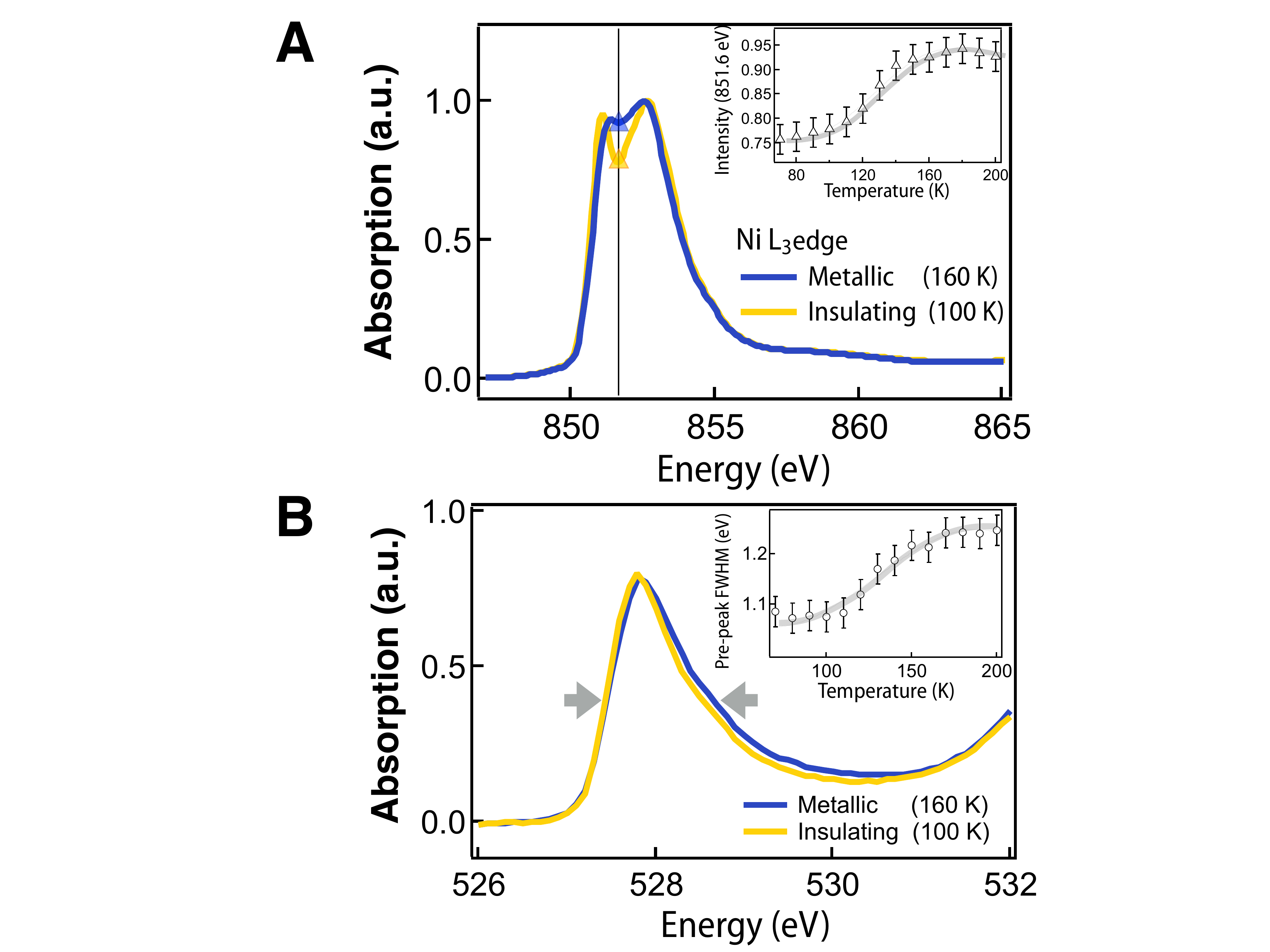}
\caption{\label{} (Color online) (\textbf{A}) XAS at the Ni L$_3$-edge for the metallic and insulating states. Inset shows the intensity between the Ni$^{3+}$ and multiplet peaks, highlighting the sudden narrowing of the peaks across the MIT. (\textbf{B}) XAS at the O K-edge for the same. Inset shows the change in the FWHM, arrows, of the O prepeak showing the bandwidth narrowing. All hatched lines are guides to the eye.}
\end{figure}

\clearpage

\begin{figure}[h!]\vspace{-0pt}
\includegraphics[width=0.8\textwidth]{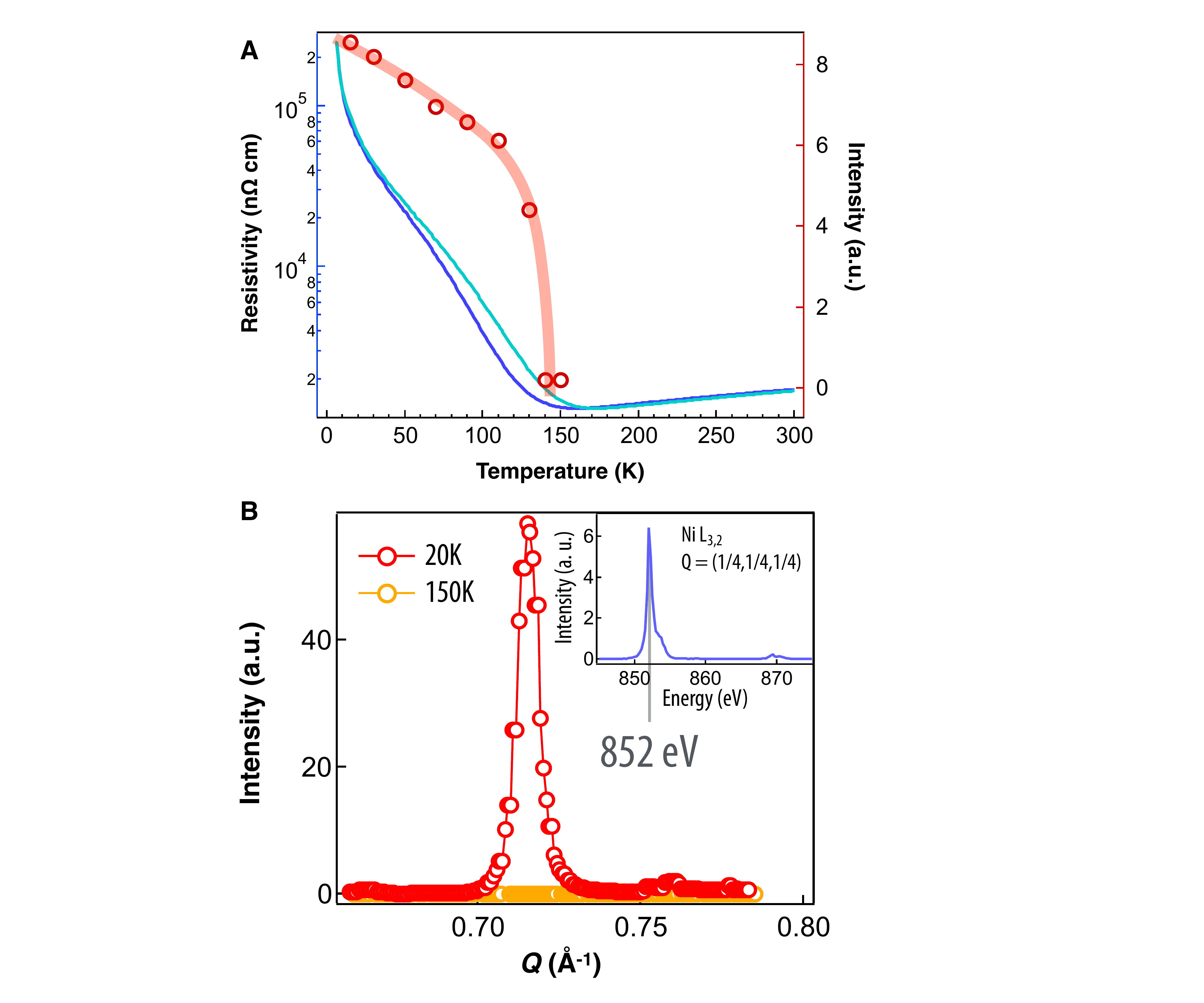}
\caption{\label{} (Color online) (\textbf{A}) Left axis: Temperature dependence DC transport for cooling (blue) and warming (cyan) cycles showing a strong hysteresis typical of the first-order MIT. Right Axis: Temperature dependence of the forbidden Bragg peak intensity corresponding to the magnetic order parameter. (\textbf{B}) Low and high temperature magnetic Bragg peak corresponding to E$^{\prime}$-type anti-ferromagnetism. The inset shows the resonant energy scan at the Ni L$_3$ and L$_2$ at the peak.}
\end{figure}

\clearpage
\begin{figure}[t!]\vspace{-0pt}
\includegraphics[width=0.8\textwidth]{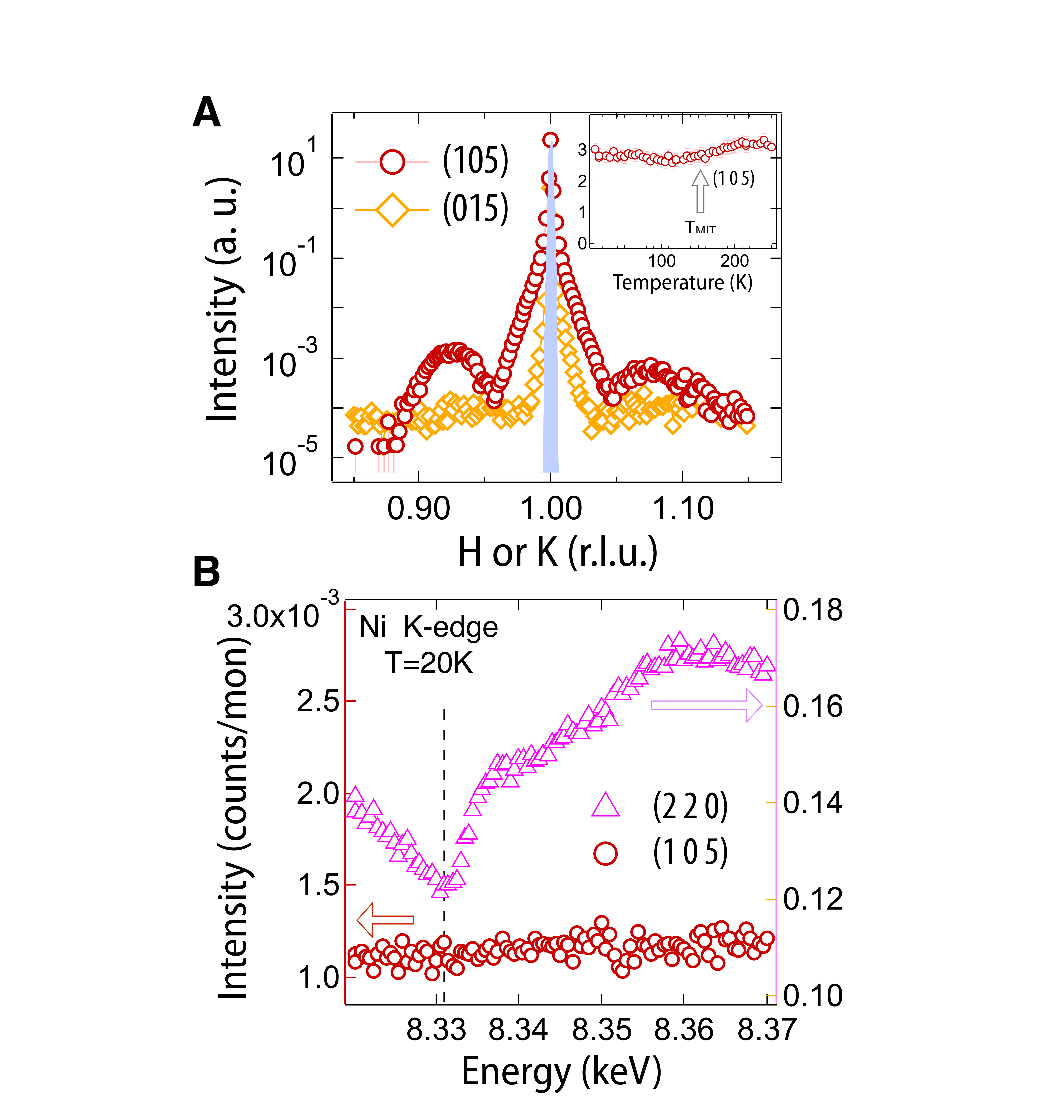}
\caption{\label{} (Color online) (\textbf{A}) Scattering around the (1 0 5)$_{or}$ and (0 1 5)$_{or}$ peaks at low temperature $\sim$ 10 $^\circ$K  (the sharp peak at 1.00 is the substrate). The inset show the measured intensity of the (1 0 5)$_{or}$ peak for several temperatures crossing the MIT. (\textbf{B}) Ni K-edge resonance scans at the (1 0 5)$_{or}$ and (2 2 0)$_{or}$ peaks.}
\end{figure}

\clearpage


\begin{thebibliography}{999}
\bibitem{Footnote1}
Distinguishing between these peaks requires single-domain films, which are found for NNO on NGO but not when grown on cubic (e.g. SrTiO$_3$) or tetragonal (e.g. SrLaAlO$_4$) substrates.
\end{thebibliography}
\end{document}